# Influence of oxygen on thermal stability of nanocrystalline aluminum studied by positron annihilation spectroscopy[*]

SONG Xian-Bao(宋先保), WANG Zhu（王 柱）[1)], TAI Peng-Fei（台鹏飞）, TIAN Feng-Shou（田丰收）, LIU Liang-Liang（刘亮亮）

College of Physics and Technology, Laboratory of nuclear solid physics, Wuhan University, Wuhan 430072, P. R. China

**Abstract:** The thermal stability of nanocrystalline (nc) Al has been studied by means of positron lifetime spectroscopy and X-ray diffraction (XRD), prepared by compacting nanoparticles under high pressures. Those nanoparticles were produced by the flow-levitation (FL) method [1] and the arc-discharge (DC) method. Effect of oxide at nanoparticle surface on structure stability was investigated especially in this experiment. The positron lifetime results reveal that vacancy clusters in grain boundaries are dominant positron traps according to the high relative intensity of the positron lifetime $\tau_2$ in both samples. The mean grain size of the sample consisting of nanoparticles with partially oxidized surfaces is almost unchanged after aging at 150℃ for 84 h, with that of the sample consisting of pure nc Al increasing. The partially oxidized surfaces of nanoparticles hinder the growth of grain when aging at 150℃, vacancy clusters related to $Al_2O_3$ need longer aging time to decompose, which implies that the oxide stabilizes the microstructure of the nanomaterials. The effect is beneficial for nc materials to keep excellent properties.

**Keywords**: Positron annihilation, Thermal stability, Nanocrystalline material, Defect, Microstructure

**PACS**: 78.70.Bj, 68.60.Dv, 81.07.Bc

## 1. Introduction

Nanocrystalline materials, with grain sizes typically smaller than 100 nm, contain a large number of interfaces, which bestow those materials advanced characters like increased strength or hardness, improved ductility or toughness, increased specific heat, lower thermal conductivity in comparison with conventional coarse grained materials [2-3]. Such materials have attracted much attention since

[*] This work was supported by the National Natural Science Foundation of China (Grant No 11275142)
Corresponding author. E-mail:wangz@whu.edu.cn

Gleiter [4] proposed them as being in a structurally different state compared with the crystalline or glassy state. These nanostructured materials are thermodynamically unstable, with a strong tendency to transform into a normal polycrystal with coarser grain size and fewer interfaces. Besides, their grains grow rapidly even at low temperatures, making it difficult for them to process and often unsuitable for usage [5-6]. Hence the stability of microstructure is essential for nanomaterials to keep those excellent performances.

In previous work, Zhou et al. [7] observed two grain-growth regimes for nc aluminum: below $T/T_m =0.78$ growth ceased at an approximate grain size of 50 nm, while at higher temperatures grain growth proceeded steadily to the submicrometer range. Claudio L. De Castro and K. Maung et al. [8-9] also observed two grain-growth regimes for nanocrystalline aluminum produced by mechanical attrition (MA). They found that grain growth occurred above the $T/T_m$ of 0.83 for samples milled in nylon or samples milled with 1 wt% diamantane in stainless steel. Mechanisms that reportedly impart thermal stability to nc microstructure prepared by MA include grain boundary pinned by impurities (solute drag), pores (pore drag), and second-phase particles (Zener drag). Other processes such as stress relaxation by grain boundary reordering, shear-migration coupling, and annealing of dislocation segments or sub-boundary remnants are possible candidates. Thus, more study is needed to identify the process responsible for the grain stability.

Positron annihilation spectroscopy (PAS) has been developed as a non-destructive technique to study open volume defects, vacancies, dislocations and vacancy clusters in solids [10]. Positrons are easily trapped by neutral and electronegactive vacancy defects which lack atomic core. Because of the reduced electron density in those defects, the trapped positron has a longer positron lifetime, which brings about different annihilation features in different defects. Therefore, the positron lifetime could be used to identify and characterizen defects [11]. In present work, we have studied the effect of oxygen on thermal stability of nc materials by monitoring defects in grain boundaries with positron annihilation technique during aging, and the influence of surface oxidation of nanoparticles on the microstructure thermal evolution of nc Al has been discussed.

2. **Experimental**

Sample A, about 13 mm in diameter and 1 mm thick(the same as sample B), was prepared by compacting Al nanoparticles produced by the flow-levitation (FL) method under a pressure of 390 MPa in vacuum, while sample B was prepared by compacting Al nanoparticles produced by the arc-discharge method (DC) under a pressure of 16 MPa in the air. Before compacted, nanoparticles in sample B was kept for 2 h at a temperature of 100 °C in the air in order to investigate the impact of surface oxidation on the microstructure thermal evolution of nc Al. Both samples were aged at 150 °C for 1–84 h in Ar gas and were cooled to room temperature while still remaining in Ar atmosphere after each aging time.

**Table 1** Characteristics of samples

| Sample | Consolidation parameters | | Nanoparticle producing method | Mean grain size | Purity of Nanoparticle |
|---|---|---|---|---|---|
| A | 390 MPa | In vacuum | FL | 50 nm | 99.99% |
| B | 16 MPa | In air | DC | 75 nm | 99.9% |

X-ray diffraction measurements were performed on a BRUKER AXS D8 ADVANCE X-ray Diffractometer, using Cu Kα radiation to estimate the mean grain size of sample A and B. Positron lifetime spectra were collected by using a conventional fast–fast time coincidence system with a time resolution of 240 ps at room temperature. The positron source employed was $^{22}$NaCl encapsulated with Ti foils, which was sandwiched between two identical samples during measuring. $1 \times 10^6$ counts were collected at least in each positron lifetime spectrum. Positron lifetime spectra were analyzed by the PATFIT program which fits the spectrum data to the sum of exponential decays after the source and background contribution being subtracted.

## 3. Results and discussion
### 3.1. XRD results and discussion

Mean grain sizes of samples measured by XRD are shown in figure 1. The mean grain size $D$ estimated using the Debye–Scherrer equation is as following:

$$FW(S) \cdot \cos(\theta) = \frac{K\lambda}{D}, \tag{1}$$

Where $FW(S)$ is the full width of the Bragg reflection peak at half maximum ($FWHM$), $\theta$ is the peak position; $K$, the value set at 0.89, is the crystallite shape factor, and wavelength $\lambda$ of Cu Kα radiation is 0.15406 nm.

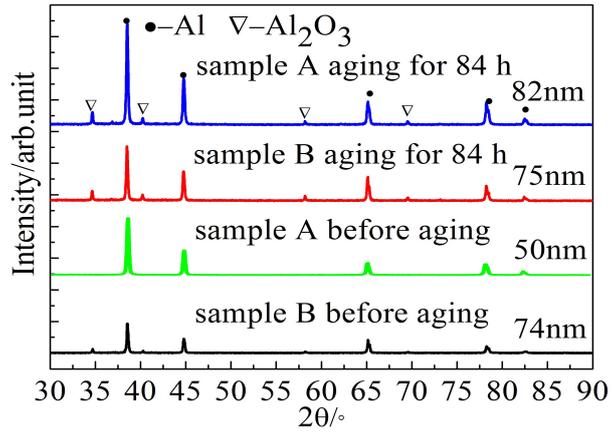

**Fig. 1.** (color online) XRD lines of samples A and B. Mean grain size is shown in right part of each XRD line.

From figure 1, we can see both samples A and B, to some extent, have been oxidized after an aging time of 84 h, and the oxides are $Al_2O_3$. For sample B, oxides come from the surface oxidation of nanoparticles as a result of being exposed in the air at 100℃ for 2 h and the sample surface during the aging process. But for sample A, oxide comes merely from the sample surface during the aging process. The mean size of sample A increases from 50 nm to 82 nm after aging for 84 hours, which indicates that the volume fraction of the disorder region in the sample decreases and more non-equilibrium atoms are recovered to equilibrium sites. But the mean grain size of sample B was almost unchanged, from 74 nm to 75 nm after aging for 84 h, suggesting that the partly oxidized surfaces of the nanoparticles retard the increase of the mean grain size in the sample.

**3.2. PLS results and discussion**

In previous studies [12-15], the following possible states of positron in nanocrystalline were suggested: (1) the free positron state (delocalized states inside the grains), (2) a trapped state related to dislocation jog in the grains or grain boundaries, (3) a trapped state related to disordered regions in grain boundaries, (4) a trapped state at monovacancy, (5) a trapped state related to a vacancy cluster containing some of the monovacancy at grain boundaries, and (6) a trapped state in large voids such as missing grains, which gives a long lifetime of nanosecond order.

In this work, all positron lifetime spectra were measured at room temperature. Three lifetime components were required to match the spectra with a variance of fit of less than 1.20 after source correction by using the PATFIT program.

As to figure 2 and figure 3, the lifetimes $\tau_1$ of both samples A and B are larger than 166 ps (the bulk positron lifetime of Al) and smaller than 253 ps (the monovacancy positron lifetime) [16] during the whole aging process. $\tau_1$ is attributed to the positron annihilating at monovacancies, delocalized states inside the grains and small defects (the size less than a monovacancy) such as dislocations and disordered regions in grain boundaries [12]. The lifetime of positron annihilating at perfect lattice in grains and small defects are smaller than that at monovacancies. The component $\tau_2$ ranging between 330 ps and 360 ps is attributed to positrons annihilating at vacancy clusters containing several monovacancies (figure 3) in grain boundaries. It is obvious that $I_2$ is much bigger than $I_1$, which suggests that vacancy clusters in grain boundaries are dominant positron traps in both samples A and B throughout the whole aging process. The positron lifetimes in $V_1$, $V_2$, $V_4$, and $V_6$ in the Al crystal (where Vn means a vacancy cluster consisting of n monovacancies) being calculated[16] are shown by horizontal lines in figure 2, 3 and 4. The longest lifetime $\tau_3$ with value about 1 to 2 ns is attributed to positronium annihilation through the pick-off process at sample surface. The intensity of $\tau_3$ is considerably small. So we will not take it into consideration in this paper.

As we known, positron average lifetime is a powerful parameter which will not be influenced by the number of components to which a positron lifetime spectrum is decomposed. From figure 2, average lifetime $\tau_A$ is evidently larger than $\tau_B$. The nanoparticles of sample B suffered from the environment of oxygen at 100℃ before compacted, which leads to partial oxidization at their surfaces. There is a large amount of interface between $Al_2O_3$ film and Al grain covered by $Al_2O_3$ film, where monovacancy, dislocation etc were easily produced, and large vacancy clusters rarely exist. Thus more monovacancies and dislocations compete with vacancy clusters for trapping positron in sample B, in comparison with sample A. This leads to a decrease in the average lifetime $\tau_B$. Another difference between $\tau_A$ and $\tau_B$ is that $\tau_A$ increases apparently, while $\tau_B$ only increases slightly along with aging time except for a sharp downward peak. In other words, the increase of average lifetime $\tau_A$ is steeper than that of $\tau_B$. The difference between $\tau_A$ and $\tau_B$ arises from the partially oxidized surfaces of nanoparticles in sample B.

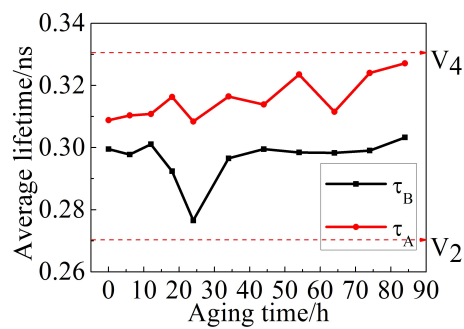

**Fig. 2.** (color online) Average lifetime for samples A and B as function of aging time at 150℃. Calculated positron lifetimes of $V_1$ and $V_4$ in Al crystal are shown by horizontal dashed lines.

In order to obtain information about the structure character during aging in details, we shall analyze the positron lifetime components and their intensities. From figure 3, $\tau_1$, $\tau_2$, $I_1$ and $I_2$ do not change at an aging time of 6 h for the sample A, from which we can infer that the defects remain stable during this process. $I_1$ increases from about 15% to 25% while $I_2$ decreases from 85% to 75%, at an aging time from 6 to 24 h. In this stage, $\tau_1$ and $\tau_2$ increase to 181 ps and 332 ps, respectively. This suggests that the concentration of vacancy clusters decreases for that unstable vacancy clusters decompose which may be caused by the release of stress. More positrons are trapped at monovacancies, the recovery of the dislocation and reorder of disorder region causes the increase in lifetime $\tau_1$. The increase of $\tau_2$ indicates that remained vacancy clusters are larger than unstable vacancy clusters in general. In fact, only slight changes are observed for the intensities $I_1$ and $I_2$ with a further aging time (over 24 h) in sample A, indicating a relatively stable defect concentration ratio formed at the temperature of 150℃. However, both lifetime $\tau_1$ and $\tau_2$ increase in a similar way, to 199 ps and 340 ps respectively, though the relative defect concentration ratio keeps stable. This indicates that the vacancies and vacancy clusters still migrate within the disordered regions and agglomerate to a larger size of vacancy cluster. In order to maintain the relative defect concentration ratio, the number of small defects decreases. This means the volume fraction of disordered regions reduced. In other words, the grain grows in sample A during the aging process. The XRD measurements also observed the grain growth in sample A from 55 nm to 82 nm after aging for 84 h.

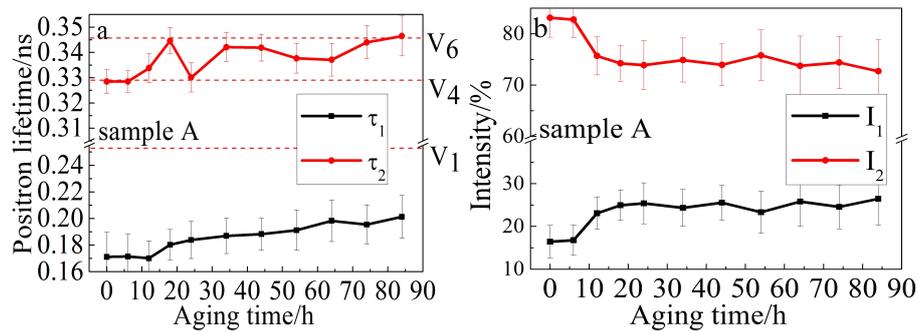

**Fig.3.** (color online) (a) Positron lifetimes and (b) intensities as function of aging time at 150℃ for sample A. Calculated positron lifetimes of $V_1$, $V_4$ and $V_6$ in Al crystal are shown in (a) by horizontal dashed lines.

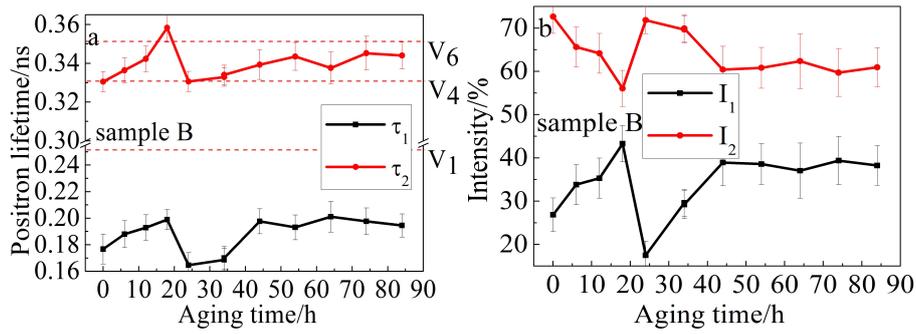

**Fig.4.** (color online) (a) Positron lifetimes and (b) intensities as function of aging time at 150℃ for sample B. Calculated positron lifetimes of $V_1$, $V_4$ and $V_6$ in Al crystal are shown in (a) by horizontal dashed lines.

From figure 4, the aging behavior of lifetime and relative intensities of sample B is similar to those of sample A at an aging time from 0 to 18 h. In this stage (0-18 h), a part of vacancy clusters decomposed which may be caused by the release of stress in sample B, which leads to a decrease of $I_2$. The increase in $\tau_1$ indicates that the recovery of dislocations in grains caused more positrons annihilation at monovacancies. Meanwhile, some of the remained vacancy clusters agglomerate and become larger, which is consistent with the increase of $\tau_2$. In the next stage (24-44 h), the changes of $\tau_1$ and $\tau_2$ are also similar to sample A, but obviously different in the behavior of $I_1$ and $I_2$. In this stage for sample B, some vacancy clusters probably related to $Al_2O_3$ undergo decomposition which may also caused by release of stress. This behaviors imply the stress was difficultly released in regions contained $Al_2O_3$. The extremely stable vacancy clusters absorb nearby monovacancies and small defects to become larger. This suggests it is hard for those vacancy clusters related to $Al_2O_3$ to decompose at previous stage or, they need longer time to decompose. The defect concentration ratio corresponding to $\tau_1$ and $\tau_2$ reaches stabilization, and positron lifetimes almost unchanged in the aging stage from 44 to 84 h. This behavior is obviously different from that of sample A, indicating sample B has not undergoing a process of grain growth. The reason is that the volume and concentration of defects decrease considerably slow due to the oxide film. So the $Al_2O_3$ hindered the growth of grain in sample B, which was proved by the XRD measurements for the size kept to a constant of 74 nm. Therefore, the oxide conduces to the structure stability of nc aluminum. Finally, after the whole process, the relative intensities of $\tau_2$ is 75% and 61% of sample A and B, respectively, indicating that vacancy clusters are difficult to be eliminated at this process.

## 4. Conclusions

Positron annihilation lifetime spectroscopy and XRD were applied to investigate the microstructure thermal evolution in nc Al, prepared by compacting two groups of nanoparticles which was produced by flow-levitation (FL) method and arc-discharge (DC) method. The average lifetime $\tau_A$ is larger than $\tau_B$, suggesting that the interface between $Al_2O_3$ film and Al matrix in sample B contains quantities dislocation and monovacancies. Vacancy clusters in grain boundaries are dominant positron traps according to the high relative intensity of the positron lifetime $\tau_2$ in both samples. The decomposition of the unstable vacancy clusters was observed in both samples at an aging time of 18 h, and vacancy clusters are difficult to be eliminated after the whole aging. For the sample B, unstable vacancy clusters related to $Al_2O_3$ need longer time to decompose. The oxide in grain boundaries stabilizes the microstructure of the nc Al and has a negative effect on grain growth, which was also observed by the XRD measurement.

## Acknowledgements

The authors would like to thank Professors H. L. Lei at China Academy of Engineering Physics for the nano particles. This work was supported by the National Natural Science Foundation of China under grant 11275142.